\documentclass[letter]{aa} 

\usepackage{graphicx}
\usepackage{wasysym}
\usepackage{txfonts}

\begin{document}

   \title{Gravitational instability and spatial regularity of the gas clouds and young stellar population in spiral arms of NGC 628}

   \titlerunning{Gravitational instability and spatial regularity in NGC 628}

   \subtitle{}

   \author{V.~S.~Kostiuk
          \inst{1}
          \and
          A.~S.~Gusev
          \inst{1}
          \and
          A.~A.~Marchuk
          \inst{2,3}
          \and
          E.~V.~Shimanovskaya
          \inst{1}}

   \institute{Sternberg Astronomical Institute, Lomonosov Moscow State 
              University, Universitetsky pr. 13, 119234 Moscow, Russia\\
              \email{valeriekostiuk@yandex.ru, gusev@sai.msu.ru}
         \and
             Central (Pulkovo) Astronomical Observatory, Russian Academy of Sciences, Pulkovskoye chaussee 65/1, 196140 St. Petersburg, Russia
         \and
             Saint Petersburg State University, 7/9 Universitetskaya nab., St. Petersburg, 199034 Russia
             }

   \date{Received March 31, 2025; accepted May 2, 2025}

  \abstract
    {There is a contradiction between the characteristic spacings in observed regular chains of star-forming regions in the spiral arms of galaxies, $\sim500$~pc, and the estimates of the wavelength of gravitational instability in them, $>1$~kpc.}
   {Calculating the scales of regularity in the grand-design galaxy NGC 628 in terms of gravitational instability, using modern high-quality observational data and comparison of them with scales of spatial regularity of the star-forming regions and molecular clouds in the spiral arms of the galaxy.}
   {We investigate two mechanisms of gravitational instability against radial and azimuthal perturbations in a multi-component disk with a finite thickness. We obtain a map of the instability wavelength distribution and compare its median value with the typical scale of observed regularity.}
   {The maps of instability parameters Q and S, which are related to radial and azimuthal perturbations respectively, show a good alignment between gravitationally unstable regions and areas of recent star formation. By analyzing the distribution of giant molecular clouds along the spiral arms of NGC 628, we found a similar regularity of about 500-600~pc that was previously observed for star-forming regions. Additionally, the distribution of the wavelength most unstable to azimuthal perturbations yielded a median value of about 700~pc, which is close to the observed scale regularity. This last finding resolved the discrepancy between the theoretically predicted and observed scales of star-forming regions' regularity.}
   {}

   \keywords{instabilities -- 
   hydrodynamics --
   ISM: kinematics and dynamics -- 
   galaxies: kinematics and dynamics -- 
   galaxies: spiral -- 
   galaxies: individual: NGC 628
               }

   \maketitle

\section{Introduction}

Turbulence compression and gravitational collapse are believed to play a key role in the creation and evolution of molecular clouds and their descendants --- the star-forming (SF) regions. The sizes and spatial distribution of gaseous (H$_2$ and H\,{\sc i}) clouds and SF~regions in galactic disks can be explained in terms of gravitational or magneto-gravitational instability \citep[see e.g.][]{elmegreen1994}.

\citet{elmegreen1983} were the first to find regular strings of H\,{\sc ii}~regions in spiral arms of galaxies. They noted the rarity of this phenomenon, merely $10\%$ of grand design galaxies studied have visually detectable regular chains of H\,{\sc ii}~regions with the typical distances between adjacent regions of 1-4 kpc in different galaxies. The regularity in the distribution of H$_2$ clouds and SF~regions implies the constancy of a number of physical parameters of the interstellar medium (ISM) and stellar population in the stellar-gaseous disk on sufficiently large scales within the disk and in a wide range of galactocentric distances, which should not often occur in typical galactic disks. Indeed, theoretical studies of the gravitational instability of gaseous, stellar-gaseous, multi-component disks \citep{Safronov1960,elmegreen1983,Jog1984,romeo2013,romeo2017,Rafikov2001}, as well as the study of the fragmentation of gas filaments, single- and multi-component spiral arms and rings \citep{inutsuka1997,mattern2018,elmegreen1994a,elmegreen1994,Inoue2018} predict regularities on scales of a few kpc for typical values of parameters of the ISM and stellar medium \citep{elmegreen1983,Marchuk2018,Inoue2021}.

However, in the last decade, regular or quasi-regular chains of SF~regions and gas clouds in spiral arms and rings have been discovered in many galaxies of various morphological types, from S0 to Scd, as well as in the Carina Arm of the Milky Way (MW) with the characteristic spacing equal to or a multiple of 350-500~pc \citep{gusev2013,elmegreen2018,Henshaw2020,gusev2020,gusev2022,proshina2022,park2023}. Recent magneto-hydrodynamic simulations in a MW-like galaxy of \citet{arora2024} also predict the formation of regular chains of gas clouds in spiral arms on scales of $\sim500$~pc in the hydrodynamic case and slightly more in the magnetic case. \citet{Meidt2023} found sub-kiloparsec characteristic distances between gas filaments and close to them turbulent Jeans lengths in four galaxies, including NGC 628, using PHANG--{\it JWST} data. Thus, recent studies have reshaped our understanding of the prevalence of the regularity phenomenon and its characteristic spatial scales.

Despite recent results of \citet{Meidt2023} and \citet{arora2024}, the question of agreement between observational data, which show a scale of $\sim500$~pc, and theoretical models (scale $>1$~kpc) remains open. The aim of this paper is to directly calculate the scales of regularity in galaxies in terms of gravitational instability, using modern high-quality surveys, which provide the necessary observational data (parameters of the ISM and stellar population, and disk kinematics).

As an object of study, we chose the well-known nearby galaxy NGC~628 (M74), which is viewed almost face-on (Fig.~\ref{fig:Q_S_lam}). It is a prototype of the grand-design galaxy, in which \citet{gusev2013} found regular chains of SF~regions in both spiral arms of the galaxy with spacing of $\approx550$~pc (for the distance 10~Mpc).

\section{Mechanisms of gravitational instability}

In this paper, we investigate two distinct mechanisms of gravitational instability that may regulate the large-scale star formation process in galaxies. Both mechanisms involve the assumption of the coexistence of gas and stellar components in a disk with a finite thickness. The first mechanism was considered in \citet{Rafikov2001} by analyzing the axisymmetric stability of a rotating disk. In order to take into account the influence of both components, rather than using the hydrodynamical approximation, as was done in \citet{Wang1994}, \citet{Rafikov2001} examined the instability process more accurately by using the collisionless Boltzmann equation for a stellar disk.  As a result, the following expression for the instability parameter was derived\footnote{Another approach is presented in \citet{romeo2013}, which is simpler and applies to realistically thick discs.}:
\begin{equation}\label{eq:Rafikov}
\frac{1}{\mathrm{Q}(\overline{k_r})}=\frac{2}{\mathrm{Q_s}}\frac{1}{\overline{k_r}}\left[1-e^{-\overline{k_r}^{2}}I_{0}(\overline{k_r^2})\right]+\frac{2}{\mathrm{Q_g}}\frac{\overline{k_r}s}{1+\overline{k_r^2}s^{2}},
\end{equation}
where $\mathrm{Q_s}=\varkappa \sigma_\mathrm{s}^r/\pi G \Sigma_{\mathrm{s}}$ and $\mathrm{Q_g}=\varkappa \sigma_\mathrm{g}^r/\pi G \Sigma_\mathrm{g}$ are classical instability parameters for stellar and gaseous disks \citep{Toomre1964}, $\overline{k_r}\equiv{k_r\,\sigma_\mathrm{s}^r}{\varkappa}^{-1}$ and $s=\sigma_\mathrm{g}^r/\sigma_\mathrm{s}^r$.
In the above equations, $\Sigma_\mathrm{s}$ and $\Sigma_\mathrm{g}$ represent the surface densities of stars and gas; $\varkappa$ and $I_{0}$ -- the epicyclic frequency and the modified Bessel function of the first kind; $\sigma_\mathrm{g,s}^r$ are the radial components of the gas and stellar velocity dispersion; $k_r$ is a wave number of radial perturbations. We assume that the disk has a finite thickness, and modify the stability criterion by adding a multiplier $F_\mathrm{g,s}(k)$ to each term in the eq.~\ref{eq:Rafikov}, where $F_\mathrm{g,s}$ is equal to $[{1 - \exp(-k h_{\mathrm{g,s}}^{z})]}/{[k h_{\mathrm{g,s}}^{z}}]$, and $h_\mathrm{g,s}^z$ are the typical vertical scales of the gas and stellar components\footnote{Note, that the commonly adopted disc thickness uses a geometrical reduction factor, rather than a physical one, which takes fully into account the vertical strucure of the disc and is thus more accurate \citet{r1,r2}.} (similarly to~\citealp{Jog1984,Elmegreen1995}).
\par 
Another mechanism examined in this paper considers the formation of giant clumps in tightly wound spirals that rotate with constant pattern speed $\Omega_p$~\citep{Inoue2018,Inoue2019}. The authors examined azimuthal perturbations in the spiral arms, whose radial slice profile was assumed to be described by a Gaussian function with a peak value $\Sigma_0$ at $\rm R_0$ and dispersion $w$. The derived instability parameter has the following form:
\begin{equation}\label{eq:Inoue}
\frac{1}{\mathrm{S}} = \frac{1}{\mathrm{S_g}} + \frac{1}{\mathrm{S_s}},\:\:\mathrm{S_{g,s}}(k_\phi) = \dfrac{(\sigma_\mathrm{g,s}^\phi k_\phi)^2 + 4 \Omega_\mathrm{p}^2}{\pi G f(k_\phi l_\mathrm{g,s})\vernal_\mathrm{g,s} F_\mathrm{g,s}(k_\phi) k_\phi^2}.
\end{equation}
In Equation \ref{eq:Inoue} $\vernal_{\rm{g},\rm{s}}=1.4 W \Sigma_0$ represents a linear mass of a certain slide of spiral with half-width $W\approx1.55w$ for gaseous and stellar components; $f(k_\phi l_\mathrm{g,s})$ is a linear combination of modified Bessel and Struve functions, as described in~\cite{Inoue2018}, depending on the wave number of the perturbation $k_\phi$ and the distance from the spiral ridge $l$. Note that in this case, $\sigma_\mathrm{g,s}^\phi$ are the azimuthal projections of the gas and stellar velocity dispersions.
\par
The formal criteria for gravitational instability in both mechanisms are that if the minimum values of the $\rm Q(k_r)$ and $\rm S(k_\phi)$ functions are less than 1, then a particular region of the disk is unstable to radial or azimuthal perturbations accordingly. Note that the value of $\lambda_{r,\phi} = 2\pi/ k_{r,\phi}$ at which $k_{r,\phi}$ leads to the minimum of these functions is considered to be the wavelength of the most unstable radial or azimuthal perturbation mode.

\section{Data and methods}

We examine the process of gravitational instability using the example of the grand-design spiral galaxy NGC~628. We assume the inclination $i$ and positional angle $\mathrm{PA}$ to be $7^\circ$ and $25^\circ$, respectively. The distance to NGC~628 is still a subject of debate. There are two alternative estimates of the distance to the galaxy: 7.2~Mpc \citep[see e.g.][]{sharina1996,dyk2006} and 9.3-10~Mpc \citep[see e.g.][]{hendry2005,olivares2010}. We adopt the distance of 10~Mpc in this paper; however, in Appendix~\ref{two_distances} we compare our obtained results for two distances, 10~Mpc and 7.2~Mpc.
\par
 We assume that the gas component of the disk consists of neutral H{\sc i} and molecular hydrogen $\rm H_2$. The map of surface density $\Sigma_{\rm HI}$ was obtained from THINGS~\citep{Walter2008} data and estimated using the standard conversion (see eq.~2 from~\citealp{Yildiz2017}). As the gas velocity dispersion, we take that of H{\sc i} obtained from THINGS. Additionally, we assume this quantity to be isotropic, so its azimuthal and radial projections are equal to the extracted $\sigma_\mathrm{g}$ value. The vertical scale of the gas disk was defined from the vertical equilibrium condition as $h_\mathrm{g} = \sigma_\mathrm{g}^2/(\pi G \Sigma_\mathrm{g})$.
\par
Since direct observations of cold molecular hydrogen in external galaxies are not trivial, it is common to estimate its abundance by observing the CO line. Using modern observational data of CO~($\mathrm{J}$=2–1) from ALMA~\citep{Leroy2021}, we estimated the surface density of $\rm H_2$ as $\Sigma_\mathrm{H_2} = \alpha^{1-0}_\mathrm{CO} R_{21}^{-1}I_\mathrm{CO(\mathrm{J}=2-1)} \cos{i}$, where $R_{21}\approx0.65$ is the ratio of the CO~($\mathrm{J}$=2–1) to CO~($\mathrm{J}$=1–0) lines~\citep{denBrok2021}. To calculate $\alpha^{1-0}_\mathrm{CO}$, we used eq.~14 from~\citet{Chiang2023}, which includes the distribution of the stellar mass and metallicity (taken from~\citealp{Williams2022}).

   \begin{figure*}
   \centering
   \includegraphics[width = 0.95\linewidth]{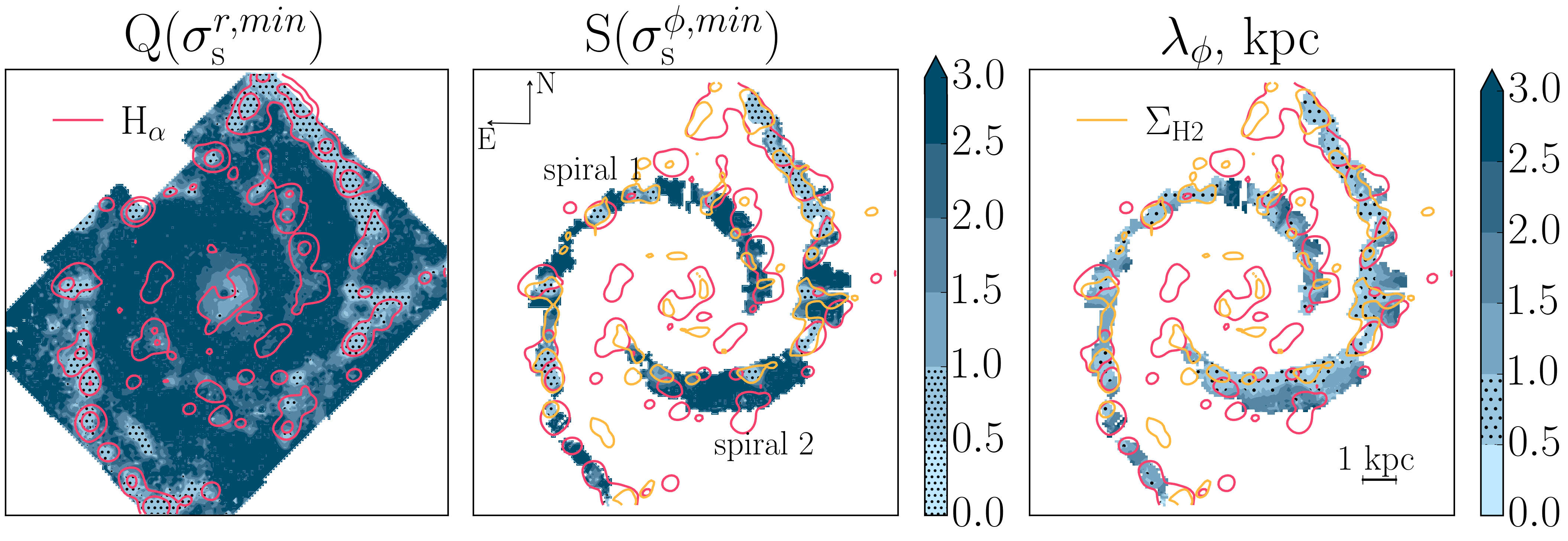}
   \caption{Left and center panels: maps of instability parameters related to radial ($\rm Q$) and azimuthal ($\rm S$) perturbations, calculated from the minimum value of the corresponding component of the stellar velocity dispersion. Right panel: the distribution of instability wavelengths relative to azimuthal perturbations $\lambda_{\phi}$. The red contours on the left indicate isolines of H$\alpha$ brightness, with values of $\log{\rm H\alpha}$ equal to $-17.2$ and $-16.5\;\rm [erg/s/cm^2]$. Red and orange contours on the middle and right panels correspond to $\log{\rm H\alpha} = -17.2\;\rm [erg/s/cm^2]$ and $\Sigma_{\rm H_2} = 30\; \rm M_{\odot}/pc^2$.}
              \label{fig:Q_S_lam}%
    \end{figure*}

Regarding the stellar component of the disk, we focused on obtaining surface density using near-infrared observations, as this relates to the old population of stars that make up the majority of the galactic stellar mass and determine its potential. We used {\it Spitzer} data at 3.6~$\mu$m and 4.5~$\mu$m bands~\citep{Sheth2010} and estimated $\Sigma_\mathrm{s}$ using eq.~8 from~\citet{Querejeta2015}. For stellar velocity dispersion, we followed a similar approach to that described in~\citet{Marchuk2017,Marchuk2018}, which allowed us to place limits on the radial ($\sigma^r_\mathrm{s}$) and azimuthal ($\sigma^\phi_\mathrm{s}$) components of velocity dispersion based on the observational line-of-sight data $\sigma_{los}$ from VLT MUSE \citep{Emsellem2022}. That way, knowing that according to observations \citep[see][]{Pinna2018,Walo-Martin2021} and numerical models \citep[][]{2003AstL...29..321S}, the $\sigma^z_\mathrm{s} / \sigma^r_\mathrm{s}$ ratio lies in the range of $0.5 \div 0.8$, we can use the epicyclic approximation: $ \sigma^\varphi/ \sigma^r = \varkappa/(2 \Omega)$~\citep{Binney2008}, and the stellar velocity ellipsoid equation to estimate, for example, the radial projection of $\sigma_\mathrm{s}$ at a certain polar angle $\varphi$:
\begin{equation*}\label{eq:dispersion}
\sigma^{r}_\mathrm{s}=\frac{\sigma_{los}}{ \sqrt{\sin^2{i}\sin^2{\varphi}+4{\varkappa}^2\Omega^{-2}\sin^2{i}\cos^2{\varphi}+m\cos^2{i}}},
\end{equation*}
where $m=0.5^2$, $0.8^2$ are taken for $\sigma^{r,max}_\mathrm{s}$, $\sigma^{r,min}_\mathrm{s}$, respectively.
The azimuthal component was derived from the radial one using the epicyclic approximation. The aforementioned epicyclic frequency and angular velocity were taken from the velocity rotation curve approximation, based on THINGS data presented in~\citet{Marchuk2018}. To estimate the vertical scale $h_\mathrm{s}$ we used the length scale taken from~\citet{Leroy2008} and the ratio between the scale length and height $q=7.3$~\citep{Kregel2002}.

To extract the linear mass of each radial slice of the spiral arms, we first determined the edges of two spirals using the method described in~\citet{Savchenko2020}. We then fitted a Gaussian function to each slice extracted from the $\Sigma_{\rm g}$ and $\Sigma_{\rm s}$ maps. As a result, we derived the surface density at the peak $\Sigma_0$ and the width of the spiral arm $W$ for each component, which allows us to obtain the linear mass $\vernal_{\rm g}$ and $\vernal_{\rm s}$. Note, by fixing the location $R_0$ where $\Sigma$ takes its peak value, we can also determine the distance to the spiral ridge $l$.  Following~\citet{Inoue2021}, the edges of each arm at a certain slice were defined as the location where $\Sigma=0.3\Sigma_0$ and the minimum $l$ cannot be less than the spatial resolution. $\Omega_p$ was estimated using corotation resonance measurements. According to~\citet{Kostiuk2024}, this galaxy may have two patterns rotating at 52 and 30~km/s/kpc.

All the aforementioned data images were rebinned to a THINGS pixel size of 1.5 arcsec/pix and cropped by the field of view defined by ALMA and MUSE images. In addition, data were also convolved with a THINGS beam size of 6.8 arcsec and 5.57 arcsec, which corresponds to a spatial resolution of approximately 150~pc at a distance of 10~Mpc.

Additionally, to the study of the distribution of SF~regions in spiral arms of NGC~628 obtained in \citet{gusev2013}, we explore the distribution of H$_2$ clouds along the spiral arms of the galaxy, based on the CO~($\mathrm{J}$=2-1) map from ALMA \citep{Leroy2021}. We used a technique developed in \citet{gusev2013} and \citet{gusev2020}. It includes along-arm photometry, finding distances $\Delta s$ between adjacent local maxima of brightness on the CO flux profiles along every arm, and analysis of their distributions, computing the Lomb-Scargle periodograms \citep{scargle1982} for the function $p(s)$, where $p(s)$ is a collection of Gaussians centered at points of local maxima of brightness on the profiles, with $\sigma$ equal to the peak positioning error. To obtain photometric profiles along spiral arms, we used the same elliptical aperture ($40\times6$~arcsec$^2$) with a minor axis along a spiral arm and a step of $1\degr$ by PA as in \citet{gusev2013}. Preliminarily, two main CO spiral arms were fitted with a logarithmic spiral with the same pitch angle, $15.7\degr$, as for stellar spiral arms in \citet{gusev2013}.

\section{Results and Discussion}

Figure~\ref{fig:Q_S_lam} (left and central panels) shows maps of the gravitational instability parameters Q and S (minimum of functions from eq.~\ref{eq:Rafikov} and~\ref{eq:Inoue}), derived from the minimum values of radial and azimuthal components of $\sigma_{\mathrm{s}}$, respectively. The maximal estimates of these parameters exceed the minimum ones by no more than a factor of 1.5, and they still indicate the same regions of instability (i.e. where Q$<$1 or S$<$1). Each image contains isolines of the H$\alpha$ brightness (MUSE/VLT) shown by red contours, which indicate regions with recent ($<10$~Myr) star formation. In most non-central areas, these contours remarkably correspond to regions gravitationally unstable to both radial and azimuthal perturbations (dotted hatching). In the central regions, star formation can be predicted by taking into account other types of perturbations that lead to an increased threshold value of the gravitational instability parameters up to 1.5-3~\citep{Morozov1985,Griv2012,Zasov2017}. Thus, we have demonstrated that for most areas in spiral arms, the parameter Q corresponds to an unstable regime, implying the ability of the disk to fragment. The S map also shows gravitational instability in spirals, primarily in gaseous clouds filled with molecular hydrogen.
\par
Despite the similar analysis of gravitational instability for NGC 628 having been conducted previously by~\citet{Marchuk2018} and~\citet{Inoue2021}, our examination led to quite different results. For instance, compared to the instability parameter map presented in the central panel of Fig.~\ref{fig:Q_S_lam},~\citet{Inoue2021} showed spiral-arm regions with higher values of S $>1$, predicting a stable state. We believe this significant difference primarily relates to the divergent methods used for estimating certain quantities, particularly the surface density of H$_2$ and old stars and the stellar velocity dispersion. In contrast to the findings of~\citet{Marchuk2018}, this study found lower values of the Q parameter, apparently due to the use of more recent data from ALMA and MUSE, and more accurate estimations.

    \begin{figure}
   \centering
   \includegraphics[width = 0.95\linewidth]{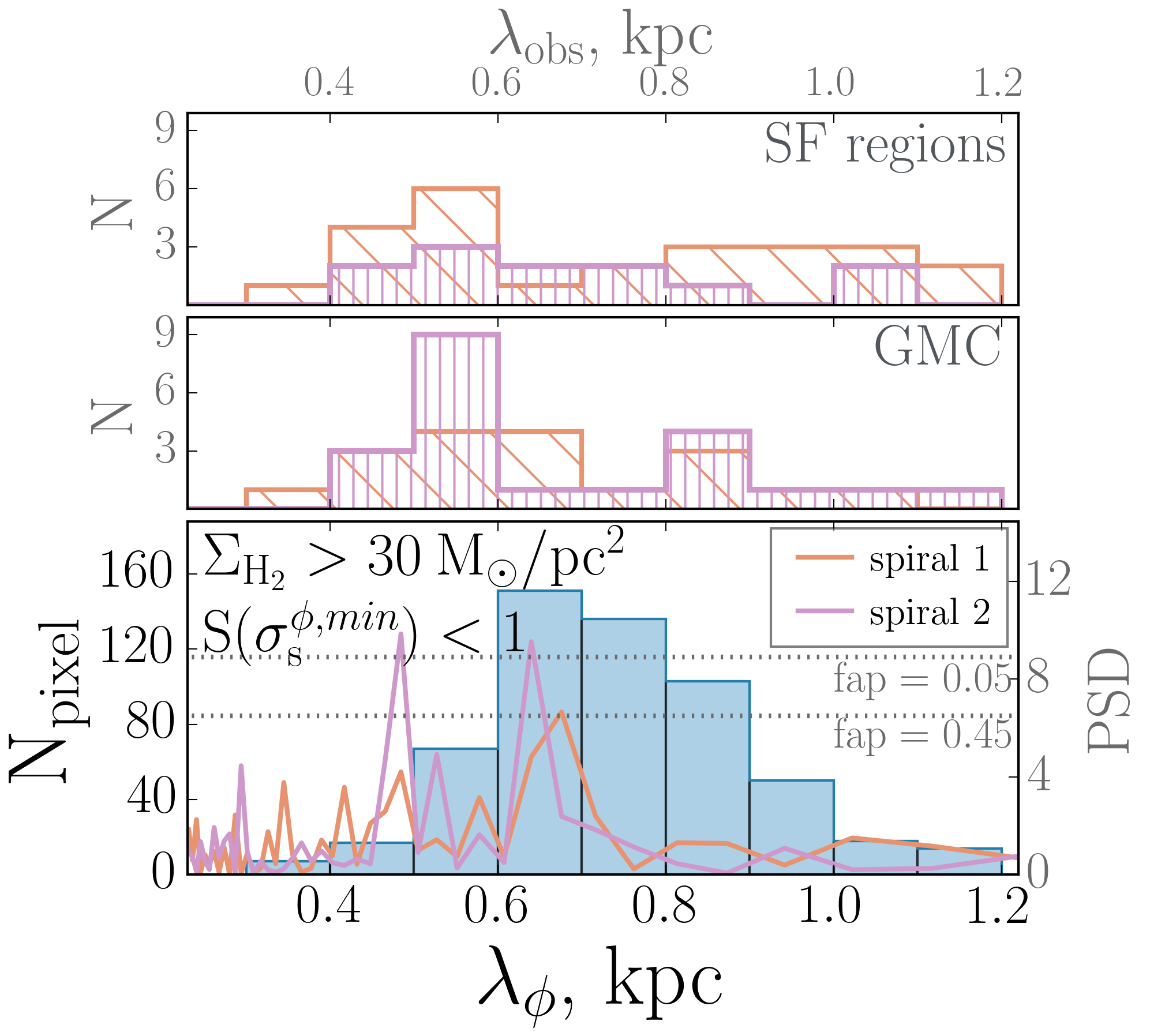}
   \caption{Top and middle panel: the histograms of the distance distributions between neighboring star-forming regions and giant molecular clouds for different spiral arms. Bottom panel: the distribution of instability wavelengths, which are shown on the right side of Fig.~\ref{fig:Q_S_lam}. The blue histogram includes only those pixels with both $\Sigma_{\rm H_2}>30\rm\: M_{\odot}/pc^2$ and $\mathrm{S}(\sigma_\mathrm{s}^{\phi,min})<1$. Solid lines represent the Fourier analysis of the GMC spacings, its range of values is shown on the right y-axis. }
    \label{fig:hist}
    \end{figure}

According to~\citet{gusev2013}, NGC~628 has regularly spaced SF~regions forming a chain-like structure along spiral arms, with a range of separations consisting $\sim$500-600~pc (assuming $D=10$~Mpc). This feature of NGC~628 is not unique, but it is rather common to other galaxies \citep[see][and references therein]{Gusev2023}. There have been several attempts to reconcile the observed spacings in the regular chains of SF~regions with theoretical calculations of the gravitational instability scales (see Introduction), although no definitive answer has been found yet. In this study, we aim to solve this puzzle by investigating the mechanism of spiral arm instability which predicts the formation of clumps. Particularly, we focus on the distribution of wavelengths of the most unstable azimuthal perturbation mode. A similar approach was performed in~\citet{Meidt2023} to explain a quasi-regular spacing of filaments by considering Jeans and Toomre wavelengths. The right panel of Fig.~\ref{fig:Q_S_lam} shows the map of $\lambda_{\phi}$ derived from wavenumber $k_\phi$ that corresponds to the minimum of the function S($k_\phi$). This image clearly shows that for most spiral pattern areas with recent star formation (red contour), wavelength of instability lies within $0.5\div1$~kpc (dotted hatches) that is consistent with the regularity spacing found in~\citet{gusev2013}. Besides, the magnitude of $\lambda_\phi$, related to the instability regions, does not vary significantly along the spiral arms, supporting the existence of regularity.

In addition to the separation between SF~regions obtained in~\citet{gusev2013}, we analyzed the distribution of local maxima of CO brightness indicating the location of giant molecular clouds (GMC). The middle panel of Fig.~\ref{fig:hist} shows two peaks, similar for both arms. The main peak contains two-thirds of separations in the range from 420 to 720~pc with a mean $\lambda_\mathrm{obs}^\mathrm{GMC}=560\pm100$~pc in both arms. The secondary peak contains the remaining one-third of the separations in the range of 840-1020~pc with a $\lambda_\mathrm{obs}^\mathrm{GMC}=900\pm80$~pc. The characteristic separations of the main peak for H$_2$ clouds are in good agreement with the results presented in the top panel for SF~regions obtained by \citet{gusev2013}, $\lambda_\mathrm{obs}^\mathrm{SF}\approx550$~pc (adopted for distance 10~Mpc). The Fourier analysis data support the presence of the spatial regularity of local maxima of CO brightness on the same scales (see solid lines at the bottom panel of Fig.~\ref{fig:hist}). The periodograms have noticeable peaks at 680~pc with a false-alarm probability (FAP)~$<45\%$ for spiral arm~1 and 480, 640~pc with FAP~$<5\%$ for arm~2.  Despite the fact that the peak for spiral arm~1 has a high FAP~$\approx45\%$ and is possibly false, the results of the Fourier analysis rather support the estimation of characteristic separations of local maxima of brightness in the arms found based on their distributions for at least one of the arms.

   \begin{figure}
   \centering
   \includegraphics[width = 0.85\linewidth]{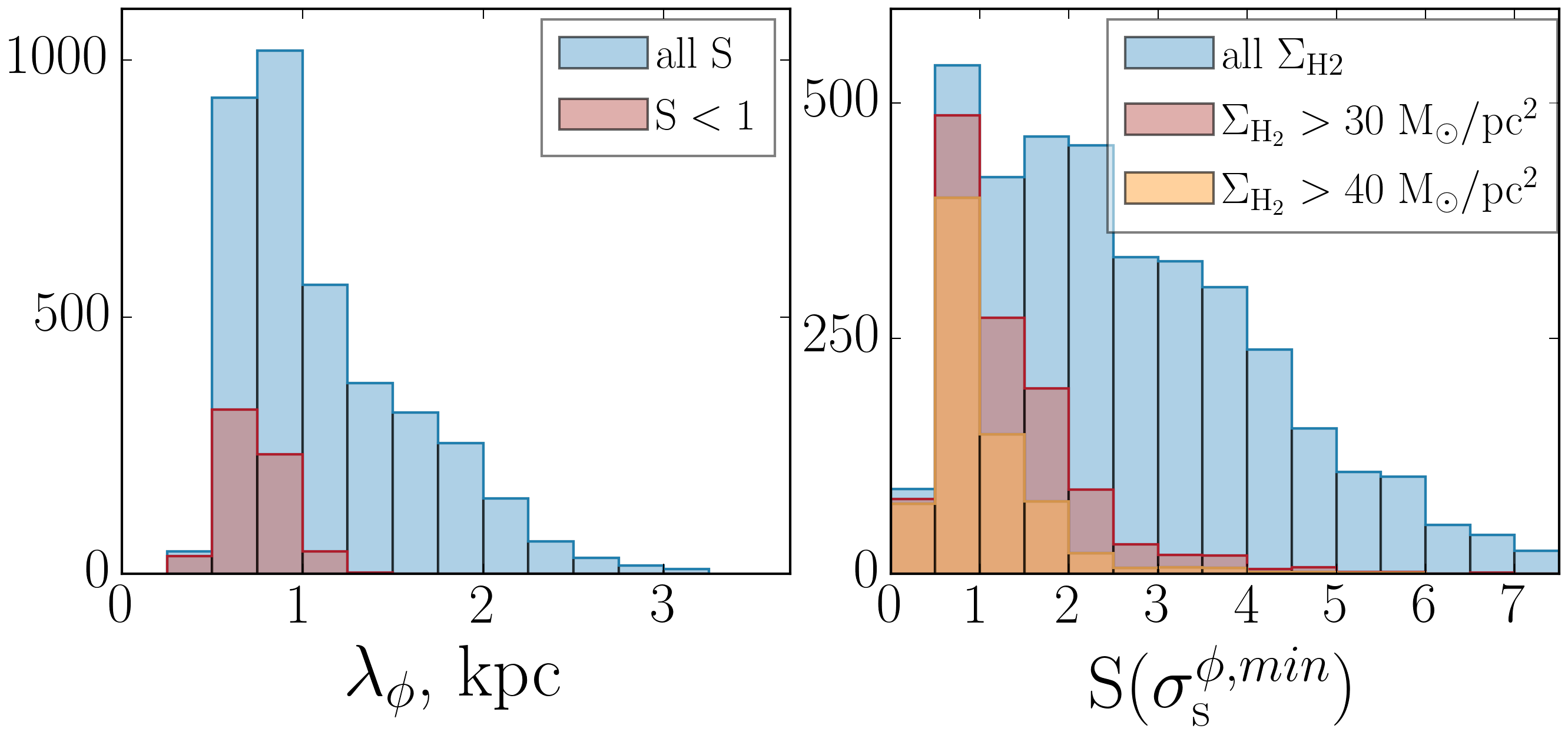}
      \caption{Left panel: the distribution of wavelength $\lambda_{\phi}$ for all regions (in blue) and for those with $\mathrm{S} < 1$ (in red). The right panel: the distribution of the instability parameter S for all pixels (in blue), and for those regions with H$_2$ surface densities greater than 30~M$_{\odot}/$pc$^2$ (in red) and 40~M$_{\odot}/$pc$^2$ (in yellow).}
         \label{fig:S_lam_distr}
   \end{figure}

   \begin{figure}
   \centering
   \includegraphics[width = 0.85\linewidth]{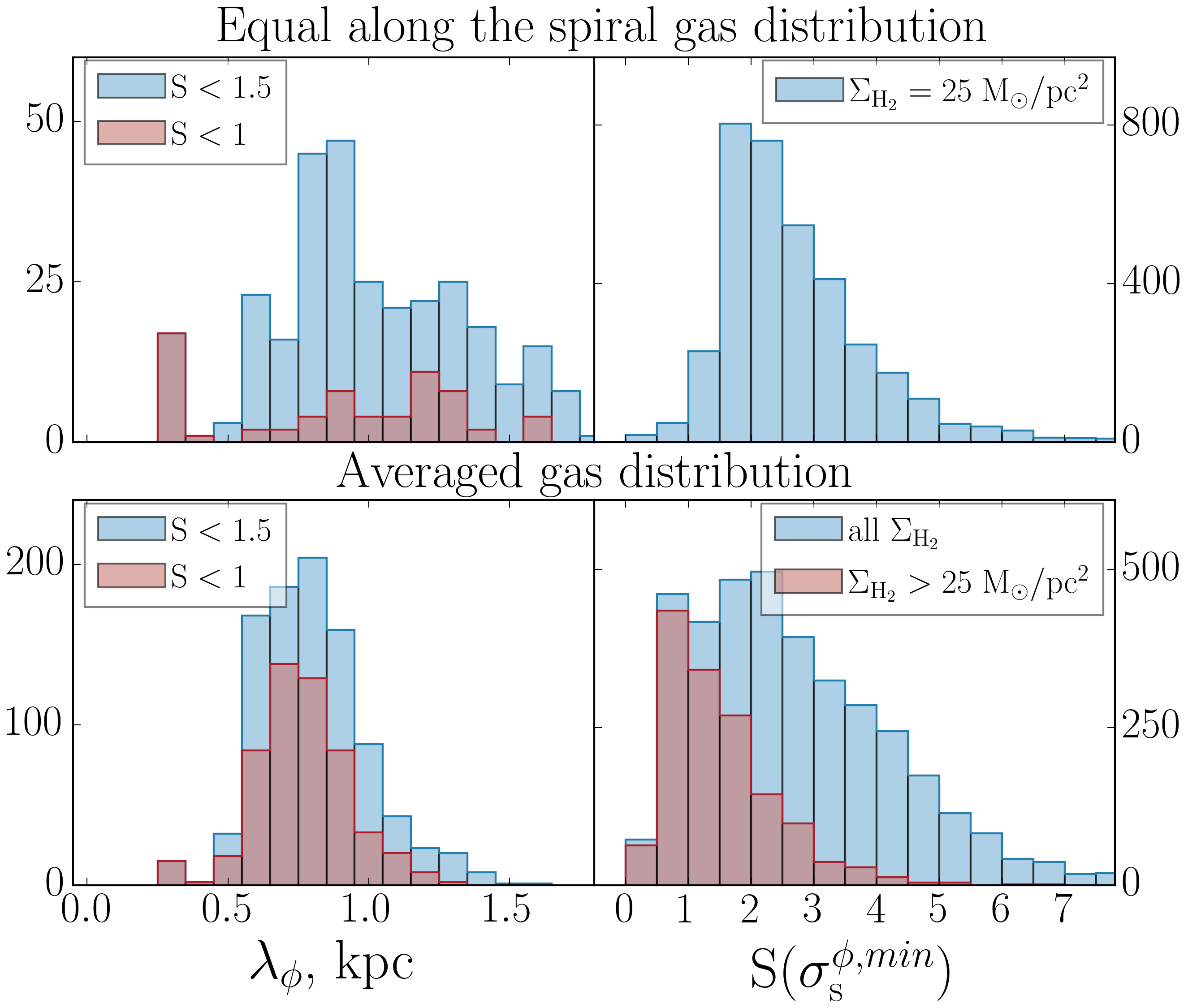}
      \caption{The same histograms as in Fig.~\ref{fig:S_lam_distr}, calculated for a gas density that is uniformly distributed along the spiral crest (top) and for an averaged gas map with a smoothing length of 300 pc (bottom).}
         \label{fig:two_cases}
   \end{figure}

The bottom panel of Fig.~\ref{fig:hist} shows the distribution of wavelengths for azimuthally unstable regions. In addition, we considered areas where the H$_2$ surface density is greater than 30~M$_{\odot}$/pc$^2$, since according to Fig.~\ref{fig:S_lam_distr}, at this threshold, the median value of the instability parameter S becomes near 1. The maximum of the $\lambda_\phi$ distribution is approximately 700~pc (the median is 720~pc) which is close to the characteristic separations of the GMCs in both arms found using Fourier analysis. These values are slightly larger than the typical separations obtained from histogram analysis. Note that the simulations of~\citet{arora2024} yielded similar results: they predicted the formation of regular chains of SF~regions in spiral arms on scales of $\sim$500~pc in the hydro case (without a magnetic field) and $\sim$650~pc in the magnetic case. However, they also found a lower median characteristic wavelength of the over-densities along spiral arms, with $\sim$730~pc in the hydro case and $\sim$1.0~kpc in the magnetic one using Fourier analysis methods.

It is worth noting that the use of current gas and other parameter distributions is not entirely correct to examine the gravitational instability regime that led to the observable distribution of star formation. Although it is nearly impossible to restore the exact distribution of matter and its kinematics prior to the collapse of the observed GMCs, we can consider two boundary cases that approximate conditions at those times. The first approach assumes the uniform distribution of gas by mass along the crest of the spiral arms. According to numerical simulations by~\citet{Inoue2018} and~\citet{arora2024}, this condition precedes the fragmentation into regularly spaced clamps that occurs within 200-300~Myr.
The second way to approximate the distribution of gas prior to clouds collapse is to smooth and average the current maps of gas density with the smoothing scale equal to the product of the free-fall collapse time and the typical velocity dispersion of gas. We assume a velocity dispersion of 10 km/s and a collapse time of 30 Myr \citep[the upper limit of times taken from][]{Chevance2020,Sun2022}; therefore, the smoothing scale is about 300 pc. For other parameters, we assume that their distributions undergo negligible changes over such short timescales. According to Fig.~\ref{fig:two_cases}, the case with an equal gas distribution demonstrates stability (S$\gg$1) almost everywhere; however, for regions with a near-unstable regime, typical wavelengths are around 1 kpc or less. In contrast, the second approximation (bottom panel) shows almost the same $\lambda_\phi$ histogram, as in Fig.~\ref{fig:hist} (bottom), with a slightly increased median value. The fact that calculated for different conditions the characteristic instability wavelength remains within the same limits can also be proven by the similar scales of regularity found separately for GMCs and young stellar clusters (see Fig.~\ref{fig:hist}).

To summarize, we have demonstrated that both instability mechanisms considered in this study can regulate the large-scale star formation process. Moreover, we were able to match the observed scale of regularity in SF regions and GMCs distributions with the theoretically predicted instability wavelength. As mentioned above, the typical regularity spacing obtained in previous investigations was a few kiloparsecs, which was significantly greater than the observable one. We believe this discrepancy might occur due to the greater magnitude of the instability parameters revealed in those studies, since according to Fig.~\ref{fig:S_lam_distr}, the wavelengths of scales less than a kiloparsec mainly correspond to an unstable regime, where the instability parameter S$<$1.

\section{Conclusions}
We examined the parameters of gravitational instability in NGC~628, referring to radial and azimuthal perturbations, including both gas and stellar components in a disk with finite thickness. A similar analysis has already been conducted for this galaxy; however, the use of recent data from MUSE and ALMA, as well as more accurate estimations of certain quantities, allows us to demonstrate a good agreement between theoretical predictions for the star formation locations and the observable ones. Besides, based on the mechanism assuming spiral arm fragmentation due to azimuthal perturbation, we obtained the wavelength of instability and compared it with the scale of SF~regions' spacing. Our conclusions are summarized as follows:
   \begin{enumerate}
      \item We demonstrated that locations with recent star formation traced by H$\alpha$ are in good agreement with the locations where parameter Q$<$1 implying gravitational instability against axisymmetric perturbations. 
      \item The map of the instability parameter S, related to azimuthal perturbations, also shows regions with unstable regime that are consistent with the locations of gaseous clouds and star-forming regions.
      \item We showed that the spatial distribution of GMCs along the spiral arms follows a regular pattern, with a characteristic scale of 500-600~pc that matches that of SF~regions.
      \item We considered the instability wavelength distribution and found that its median value ($\approx700$~pc) was close to the observed scale of regularity.
   \end{enumerate}

Thus, using the example of NGC 628, we apparently managed to remove the contradiction between the observed characteristic spacings in regular chains of SF~regions and GMCs, $\sim0.5-0.6$~kpc, and earlier calculations of the gravitational instability wavelength, $>1$~kpc.

\begin{acknowledgements}
We are grateful to the referee for his/her constructive comments. The authors would like to thank Natalia Ya. Sontikova for her useful comments and valuable feedback on the interpretation of the findings, which improved this study. The study was conducted under the state assignment of Lomonosov State Moscow University. VK thanks the Theoretical Physics and Mathematics Advancement Foundation ''BASIS'' grant number 23-2-2-6-1. 
\end{acknowledgements}

%  \bibliographystyle{aa} 
%  \bibliography{aa54886-25} 

\begin{thebibliography}{101}

\bibitem[Arora et al.(2024)]{arora2024}
Arora, R., Federrath, C., Banerjee, R., \& K{\"o}rtgen, B. 2024, A\&A, 687, A276

\bibitem[Binney \& Tremaine(2008)]{Binney2008}
Binney, J. \& Tremaine, S. 2008, Galactic Dynamics: Second Edition

\bibitem[Chevance et al.(2020)]{Chevance2020}
Chevance, M., Kruijssen, J. M. D., Hygate, A. P. S., et al. 2020, MNRAS, 493, 2872

\bibitem[Chiang et al.(2023)]{Chiang2023}
Chiang, I.-D., Hirashita, H., Chastenet, J., et al. 2023, MNRAS, 520, 5506

\bibitem[den Brok et al.(2021)]{denBrok2021}
den Brok, J. S., Chatzigiannakis, D., Bigiel, F., et al. 2021, MNRAS, 504, 3221

\bibitem[Elmegreen(1994a)]{elmegreen1994a}
Elmegreen, B. G. 1994a, ApJ, 425, L73

\bibitem[Elmegreen(1994b)]{elmegreen1994}
Elmegreen, B. G. 1994b, ApJ, 433, 39

\bibitem[Elmegreen(1995)]{Elmegreen1995}
Elmegreen, B. G. 1995, MNRAS, 275, 944

\bibitem[Elmegreen \& Elmegreen(1983)]{elmegreen1983}
Elmegreen, B. G. \& Elmegreen, D. M. 1983, MNRAS, 203, 31

\bibitem[Elmegreen et al.(2018)]{elmegreen2018}
Elmegreen, B. G., Elmegreen, D. M., \& Efremov, Y. N. 2018, ApJ, 863, 59

\bibitem[Emsellem et al.(2022)]{Emsellem2022}
Emsellem, E., Schinnerer, E., Santoro, F., et al. 2022, A\&A, 659, A191

\bibitem[Griv \& Gedalin(2012)]{Griv2012}
Griv, E. \& Gedalin, M. 2012, MNRAS, 422, 600

\bibitem[Gusev(2023)]{Gusev2023}
Gusev, A. S. 2023, Astron. Rep., 67, 458

\bibitem[Gusev \& Efremov(2013)]{gusev2013}
Gusev, A. S. \& Efremov, Y. N. 2013, MNRAS, 434, 313

\bibitem[Gusev \& Shimanovskaya(2020)]{gusev2020}
Gusev, A. S. \& Shimanovskaya, E. V. 2020, A\&A, 640, L7

\bibitem[Gusev et al.(2022)]{gusev2022}
Gusev, A. S., Shimanovskaya, E. V., \& Zaitseva, N. A. 2022, MNRAS, 514, 3953

\bibitem[Hendry et al.(2005)]{hendry2005}
Hendry, M. A., Smartt, S. J., Maund, J. R., et al. 2005, MNRAS, 359, 906

\bibitem[Henshaw et al.(2020)]{Henshaw2020}
Henshaw, J. D., Kruijssen, J. M. D., Longmore, S. N., et al. 2020, Nature Astronomy, 4, 1064

\bibitem[Inoue et al.(2021)]{Inoue2021}
Inoue, S., Takagi, T., Miyazaki, A., et al. 2021, MNRAS, 506, 84

\bibitem[Inoue \& Yoshida(2018)]{Inoue2018}
Inoue, S. \& Yoshida, N. 2018, MNRAS, 474, 3466

\bibitem[Inoue \& Yoshida(2019)]{Inoue2019}
Inoue, S. \& Yoshida, N. 2019, MNRAS, 485, 3024

\bibitem[Inutsuka \& Miyama(1997)]{inutsuka1997}
Inutsuka, S.-i. \& Miyama, S. M. 1997, ApJ, 480, 681

\bibitem[Jog \& Solomon(1984)]{Jog1984}
Jog, C. J. \& Solomon, P. M. 1984, ApJ, 276, 114

\bibitem[Kostiuk et al.(2024)]{Kostiuk2024}
Kostiuk, V. S., Marchuk, A. A., \& Gusev, A. S. 2024, Res. Astron. Astrophys., 24, 075007

\bibitem[Kregel et al.(2002)]{Kregel2002}
Kregel, M., van der Kruit, P. C., \& de Grijs, R. 2002, MNRAS, 334, 646

\bibitem[Leroy et al.(2021)]{Leroy2021}
Leroy, A. K., Schinnerer, E., Hughes, A., et al. 2021, ApJS, 257, 43

\bibitem[Leroy et al.(2008)]{Leroy2008}
Leroy, A. K., Walter, F., Brinks, E., et al. 2008, AJ, 136, 2782

\bibitem[Marchuk(2018)]{Marchuk2018}
Marchuk, A. A. 2018, MNRAS, 476, 3591

\bibitem[Marchuk \& Sotnikova(2017)]{Marchuk2017}
Marchuk, A. A. \& Sotnikova, N. Y. 2017, MNRAS, 465, 4956

\bibitem[Mattern et al.(2018)]{mattern2018}
Mattern, M., Kainulainen, J., Zhang, M., \& Beuther, H. 2018, A\&A, 616, A78

\bibitem[Meidt et al.(2023)]{Meidt2023}
Meidt, S. E., Rosolowsky, E., Sun, J., et al. 2023, ApJ, 944, L18

\bibitem[Morozov(1985)]{Morozov1985}
Morozov, A. G. 1985, Soviet Ast., 29, 120

\bibitem[Olivares et al.(2010)]{olivares2010}
Olivares E. F., Hamuy, M., Pignata, G., et al. 2010, ApJ, 715, 833

\bibitem[Park et al.(2023)]{park2023}
Park, G., Koo, B.-C., Kim, K.-T., \& Elmegreen, B. 2023, ApJ, 955, 59

\bibitem[Pinna et al.(2018)]{Pinna2018}
Pinna, F., Falc{\'o}n-Barroso, J., Martig, M., et al. 2018, MNRAS, 475, 2697

\bibitem[Proshina et al.(2022)]{proshina2022}
Proshina, I. S., Moiseev, A. V., \& Sil'chenko, O. K. 2022, Astron. Lett., 48, 139

\bibitem[Querejeta et al.(2015)]{Querejeta2015}
Querejeta, M., Meidt, S. E., Schinnerer, E., et al. 2015, ApJS, 219, 5

\bibitem[Rafikov(2001)]{Rafikov2001}
Rafikov, R. R. 2001, MNRAS, 323, 445

\bibitem[Romeo(1992)[]{r1}
Romeo, A. B. 1992, MNRAS, 256, 307

\bibitem[Romeo(1994)[]{r2}
Romeo, A. B. 1994, A\&A, 286, 799

\bibitem[Romeo \& Falstad(2013)[]{romeo2013}
Romeo, A. B. \& Falstad, N. 2013, MNRAS, 433, 1389

\bibitem[Romeo \& Mogotsi(2017)]{romeo2017}
Romeo, A. B. \& Mogotsi, K. M. 2017, MNRAS, 469, 286

\bibitem[Safronov(1960)]{Safronov1960}
Safronov, V. S. 1960, Annales d’Astrophysique, 23, 979

\bibitem[Savchenko et al.(2020)]{Savchenko2020}
Savchenko, S., Marchuk, A., Mosenkov, A., \& Grishunin, K. 2020, MNRAS, 493, 390

\bibitem[Scargle(1982)]{scargle1982}
Scargle, J. D. 1982, ApJ, 263, 835

\bibitem[Sharina et al.(1996)]{sharina1996}
Sharina, M. E., Karachentsev, I. D., \& Tikhonov, N. A. 1996, A\&AS, 119, 499

\bibitem[Sheth et al.(2010)]{Sheth2010}
Sheth, K., Regan, M., Hinz, J. L., et al. 2010, PASP, 122, 1397

\bibitem[Sotnikova \& Rodionov(2003)]{2003AstL...29..321S}
Sotnikova, N. Y. \& Rodionov, S. A. 2003, Astron. Lett., 29, 321

\bibitem[Sun et al.(2022)]{Sun2022}
Sun, J., Leroy, A. K., Rosolowsky, E., et al. 2022, AJ, 164, 43

\bibitem[Toomre(1964)]{Toomre1964}
Toomre, A. 1964, ApJ, 139, 1217

\bibitem[Van Dyk et al.(2006)]{dyk2006}
Van Dyk, S. D., Li, W., \& Filippenko, A. V. 2006, PASP, 118, 351

\bibitem[Walo-Mart{\'\i}n et al.(2021)]{Walo-Martin2021}
Walo-Mart{\'\i}n, D., P{\'e}rez, I., Grand, R. J. J., et al. 2021, MNRAS, 506, 1801

\bibitem[Walter et al.(2008)]{Walter2008}
Walter, F., Brinks, E., de Blok, W. J. G., et al. 2008, AJ, 136, 2563

\bibitem[Wang \& Silk(1994)]{Wang1994}
Wang, B. \& Silk, J. 1994, ApJ, 427, 759

\bibitem[Williams et al.(2022)]{Williams2022}
Williams, T. G., Kreckel, K., Belfiore, F., et al. 2022, MNRAS, 509, 1303

\bibitem[Y{\i}ld{\i}ız et al.(2017)]{Yildiz2017}
Y{\i}ld{\i}ız, M. K., Serra, P., Peletier, R. F., Oosterloo, T. A., \& Duc, P.-A. 2017, MNRAS, 464, 329 440

\bibitem[Zasov \& Zaitseva(2017)]{Zasov2017}
Zasov, A. V. \& Zaitseva, N. A. 2017, Astron. Lett., 43, 439 

\end{thebibliography}

\appendix
\section{Instability parameters in the case of alternative distance}
\label{two_distances}
\begin{figure}
   \centering
   \includegraphics[width = 0.92\linewidth]{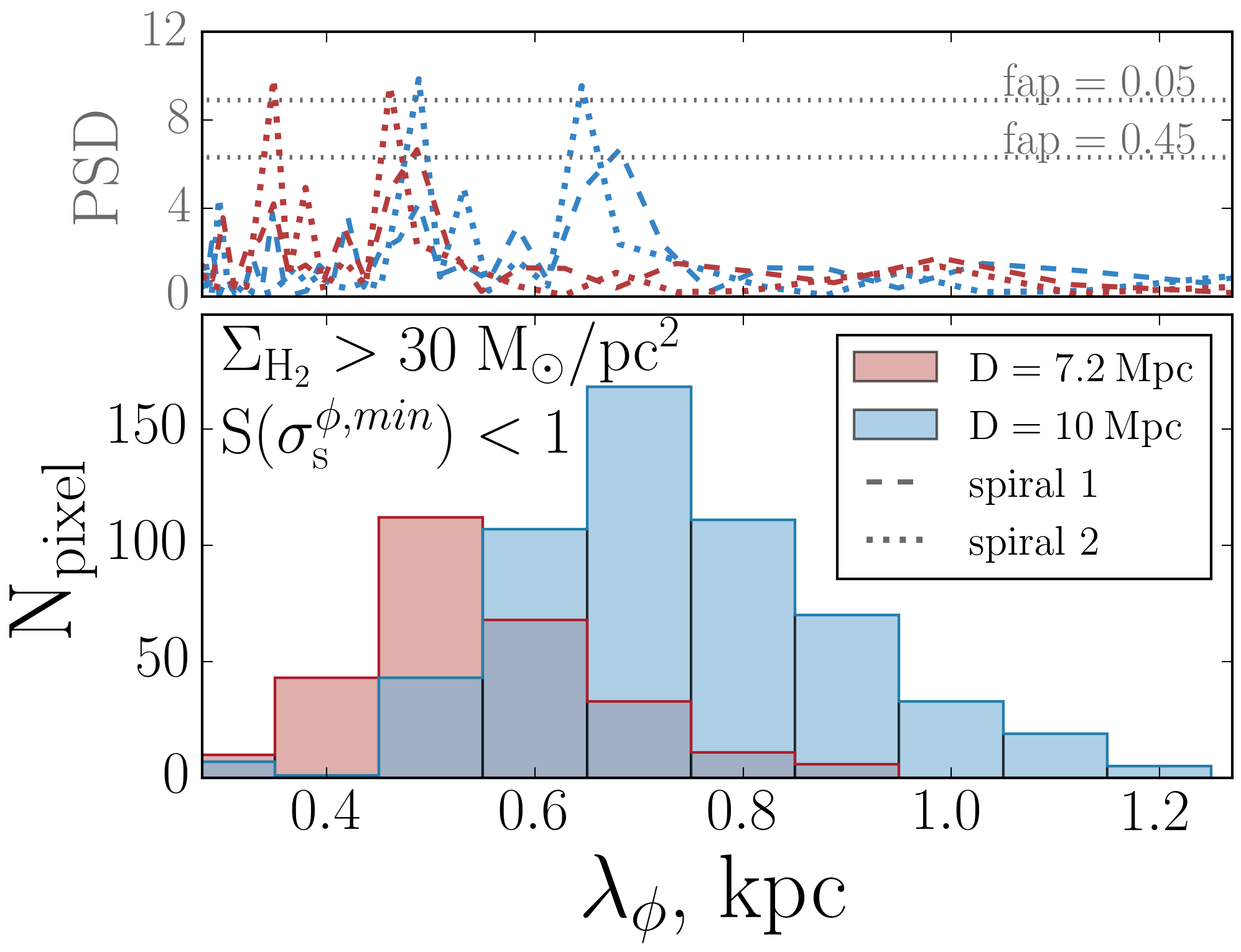}
      \caption{Top panel: the same periodograms as in Fig.~\ref{fig:hist}. The dashed and dotted lines indicate different spiral arms. Bottom panel: the same distribution as at the bottom of Fig.~\ref{fig:hist}. Colors of the lines and histograms indicate the distance to the galaxy used in the calculations, with red representing $7.2$ Mpc and blue representing $10$ Mpc. }
         \label{two_dist}
   \end{figure}

We examined the distribution of instability wavelengths and regular spacing of GMCs for a distance of 7.2~Mpc (see Fig.~\ref{two_dist}, red histogram and red lines). As can be seen from this figure, both theoretical and observed $\lambda$ decrease with distance. Indeed, the distance between GMCs is lower for $D=7.2$~Mpc proportionally to the ratio of both distances. Both the instability wavelength's magnitude and the value of the instability parameter itself do not directly depend on distance, since the only quantities that are linked to the distance are the angular speed of the pattern and the typical vertical scale. Nevertheless, the median of the $\lambda_\phi$ distribution (530 pc) matches the characteristic separation seen in periodograms (shown on the top), which suggests that the revealed result is not dependent on distance. However, it is worth noting the decreased number of pixels in the red histogram, which indicates that lower distance assumptions lead to stabilization of the spiral arms matter against azimuthal perturbations. This can be explained by the fact that, for lower distances, the spiral pattern is rotating faster, apparently preventing the compression of gas clouds implying star formation. In addition, the higher pattern speed can cause gravitationally bound clouds to split into smaller parts, leading to decreasing wavelength values.

\end{document}